# What do halo CMEs tell us about solar cycle 25?

Nat Gopalswamy[1], Grzegorz Michalek[2], Seiji Yashiro[1,3], Pertti Mäkelä[1,3], Sachiko Akiyama[1,3], and Hong Xie[1,3]

[1]NASA Goddard Space Flight Center, Greenbelt, MD 20771

[2]Astronomical Observatory of the Jagiellonian University, Krakow, Poland

[3]The Catholic University of America, Washington DC 20624

Abstract

It is known that the weak state of the heliosphere due to diminished solar activity in cycle 24 back-reacted on coronal mass ejections (CMEs) to make them appear wider for a given speed. One of the consequences of the weak state of the heliosphere is that more CMEs appear as halo CMEs (HCMEs), and halos are formed at shorter heliocentric distances. Current predictions for the strength of solar cycle (SC) 25 range from half to twice the strength of SC 24. We compare the HCME occurrence rate and other properties during the rise phase of cycles 23, 24, and 25 to weigh in on the strength of SC 25. We find that HCME and solar wind properties in SC 25 are intermediate between SCs 23 and 24, but closer to SC 24. The HCME occurrence rate, normalized to the sunspot number, is higher in SCs 24 and 25 than in SC 23. The solar wind total pressure in SC 25 is ~35% smaller than that in SC 23. Furthermore, the occurrence rates of high-energy solar energetic particle events and intense geomagnetic storms are well below the corresponding values in SC 23, but similar to those in SC 24. We conclude that cycle 25 is likely to be similar to or slightly stronger than cycle 24, in agreement with polar-field precursor methods for cycle 25 prediction.

## 1. Introduction

Halo coronal mass ejections (CMEs) were first identified in the Solwind coronagraph images of the Solwind coronagraph on board the P78-1 spacecraft (Howard et al. 1982; 1985). The name reflects the observational fact that a brightness enhancement is observed all around the occulting disk of the observing coronagraph (Gopalswamy et al. 2003; Gopalswamy et al. 2007; Gopalswamy et al. 2010a; Zhang et al. 2010) and the apparent CME width is designated as 360⁰. In reality, a three-dimensional CME of finite angular width is projected on the sky plane and hence appears as a halo. Simultaneous observations from coronagraphs on board the Solar and Heliospheric Observatory (SOHO) and the Solar Terrestrial Relations Observatory (STEREO) when the spacecraft were in quadrature revealed that halo CMEs (HCMEs) are regular CMEs that are fast and wide on the average (Gopalswamy et al. 2012a; 2013; Makela et al. 2016 ). This is also revealed by the fact that the fraction of halos is high in energetic CME populations, e.g., ~58% in CMEs causing low frequency type II radio bursts (Gopalswamy et al. 2019a), ~67% in CMEs causing intense geomagnetic storms (Dst < -100 nT), ~80% in CMEs associated with large solar energetic particle (SEP) events (Gopalswamy 2022), and ~100% in CMEs associated with sustained gamma-ray emission (Gopalswamy et al. 2018; 2019b). These fractions are an order of magnitude higher than the few percent in the general population of CMEs.

To appear as halos, CMEs need to originate close to the central meridian either in the frontside or the backside of the Sun as viewed from the coronagraph, although about 10% of halo CMEs originate from the limb. Limb halos are exceptionally fast (~1400 km/s), a few times faster than the average speed of the general population (Gopalswamy et al. 2010b; 2020a; Cid et al. 2012). Thus, a frontside halo CME observed from the Sun-Earth line has a high probability of impacting Earth and causing a geomagnetic storm (Srivastava and Venkatakrishnan 2004; Gopalswamy et al. 2007; Webb and Howard, 2012; Shen et al. 2014; Dumbovic et al. 2015; Scolini et al. 2018; Schmieder et al. 2020; Pricopi et al. 2022; Mishra and Teriaca, 2023).

The SOHO observations of CMEs in solar cycle (SC) 24 indicated that there is something peculiar about the HCME occurrence rate when compared with that in SC 23: the number of HCMEs did not drop significantly while the sunspot number (SSN) decreased by more than 40% (Gopalswamy et al. 2015a; Dagnew et al. 2020). The higher abundance of halos in SC 24 (normalized to SSN) was found to be an indicator of the weakness of that cycle attributed to the backreaction of the heliospheric state on CMEs. The reduced pressure in the heliosphere made CMEs expand more so they became halos even while originating from larger central meridian distance (CMD). Furthermore, SC 24 CMEs became halos sooner and at a lower speed than the SC 23 ones (Gopalswamy et al. 2020a,b). These results suggest that it is possible to infer the relative strength of a solar cycle based on the occurrence rate of HCMEs. Now that cycle 25 is approaching its maximum, we are in a position to compare HCMEs during the rise phase of three SCs to shed light on the strength of SC 25.

We focus on the rise phase because the daily rate of CMEs and SSN are highly correlated in the rise phase of the solar activity cycle (Gopalswamy 2022). We use the close correlation between SSN and HCME occurrence rate to assess the relative strength of SC 25. There have been a large number of predictions of the strength of SC 25 (Nandy 2021 and references therein). Precursor-based and physics-based methods indicate that SC 25 will be similar to, or slightly stronger than SC 24 (Nandy et al. 2021). On the other hand, there are other methods that predict a much stronger cycle (Han and Yin 2019; McIntosh et al. 2020; 2023). The halo CME rate thus provides an independent assessment of the cycle strength based on the fact that a weak cycle makes more halos due to the backreaction of the weakened heliosphere on CMEs (Gopalswamy et al. 2015a; Dagnew et al. 2020).

## 2. Observations

Our main objective is to compare the HCME occurrence rates corresponding to the rise phase of SCs 23-25. We know the rise phase durations of SCs 23 and 24. In SC 25, we are currently in the rise phase, but we do not know the exact start of the maximum phase. Therefore, we compare the first 37 months in each cycle. We consider all HCMEs recorded by the Large Angle and Spectrometric Coronagraph (LASCO, Brueckner et al., 1995) on board SOHO from 1996 until the end of 2022. A catalog of HCMEs is available at the CDAW Data Center (http://cdaw.gsfc.nasa.gov/CME_list/halo/halo.html, see Gopalswamy et al. 2010a for details). The catalog includes the heliographic coordinates of the associated eruptions along with the sky-plane and cone-model corrected speeds of the HCMEs. The solar sources are identified using the locations of the associated flares (from NOAA Space Weather Prediction Center), filament eruptions or coronal dimming (identified in the EUV images obtained by the Atmospheric

Imaging Assembly (AIA, Lemen et al. 2012) on board the Solar Dynamic Observatory (SDO, Pesnell et al. 2012). In SCs 24 and 25, the locations of backside halos are confirmed using EUV images obtained by STEREO's Sun Earth Connection Coronal and Heliospheric Investigation (SECCHI, Howard et al. 2008). The SSN data (v2) are from SILSO, the World Data Center – Sunspot Index and Long-term Solar Observations (http://www.sidc.be/silso/datafiles#total). We count all HCMEs in each Carrington Rotation (CR) and obtain the daily rate by dividing the total number by 27.3 days. We perform a similar averaging of the daily SSN over each CR. These parameters are then used in the correlation analysis. To assess the heliospheric state, we use OMNI data (https://omniweb.gsfc.nasa.gov/, Papitashvili and King 2020) to track the variation of various solar wind parameters. We also use the disturbance storm time (Dst) index made available by the World Data Center for Geomagnetism in Kyoto (https://wdc.kugi.kyoto-u.ac.jp/). Finally, we use the list on ground level enhancement (GLE) in SEP events from the Oulu GLE data base (https://gle.oulu.fi/).

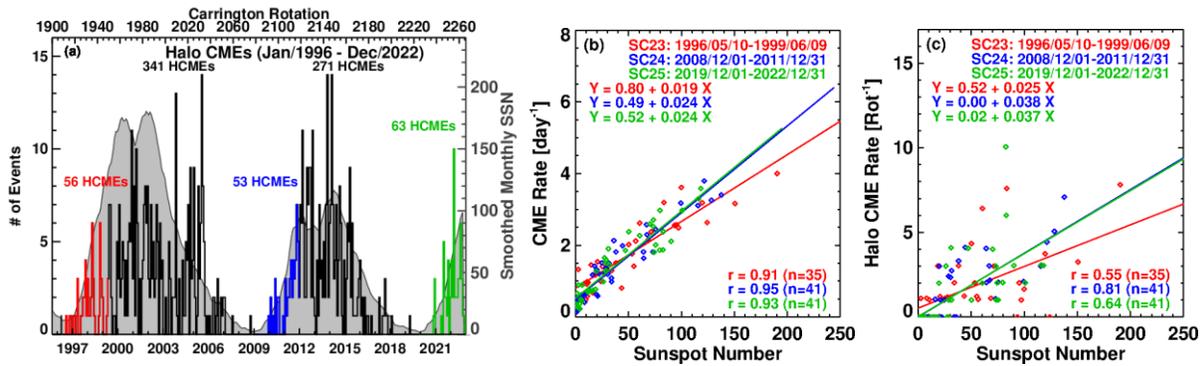

Figure 1 (a) The number of HCMEs summed over Carrington rotation periods plotted as a function of time since the beginning of 1996. The 13-month smoothed monthly SSN is shown in gray for reference. The rise phases of SCs 23, 24, and 25 are distinguished by the red, blue, and green segments of the plot. (b) Scatter plot between CME daily rate averaged over Carrington Rotation (CR) periods and the Sunspot number (v2) for the rise phase of three SCs: 23 (red), 24 (blue), and 25 (green). (c) Scatter plot between the SSN and HCME occurrence rate (per CR). The number of data points in SC 23 is smaller because we dropped 6 CRs that had <50% duty cycle due to SOHO data gap. All the correlation coefficients in the plots exceed the Pearson's critical coefficients: for 35 and 41 samples are 0.53 and 0.50, respectively with the chance-coincidence probability $<5\times10^{-4}$.

## 3. Analysis and Results

### 3.1 HCME rate and SSN

Figure 1a shows the relation between SSN and the HCME occurrence rate (number per CR) since 1996. The number of HCMEs during the first 37 months in each solar cycle is distinguished by colors: red (SC 23), blue (SC 24), and green (SC 25). The black histogram denotes the rate during the rest of SCs 23 and 24. The HCME rate in Fig. 1a shows an overall good correlation with SSN. However, there are some marked differences such as the largest bins in the declining phase of SC 23 due to certain highly CME-productive active regions. We see that the number of HCMEs during the first 37 months in each cycle is about the same: 56, 53, and 63 in SCs 23, 24, and 25, respectively. On the other hand, the daily SSN averaged over the first three years are 53.4, 33.8, and 40.5, respectively. The SSN in SCs 24 and 25 are smaller by 37% and 24% than in SC 23. In cycle 23, there was a 4.5-month data gap when SOHO was temporarily disabled. Therefore, the 56 HCMEs in SC 23 is an underestimate. If HCMEs occurred at the same rate as the average rate in the rise phase of SC 23, we estimate an additional 8 HCMEs bringing the total to ~64. The number of HCMEs normalized to SSN becomes 1.20, 1.57, and 1.56, respectively in SCs 23, 24, and 25. Clearly, the rates are similar in cycles 24 and 25 but higher than that in SC 23 by ~30%. This trend of higher abundance of HCMEs for a smaller SSN, first shown during SC 24, holds good for SC 25 as well.

As noted in the introduction, the CME daily rate is highly correlated with SSN. This is further illustrated in Fig. 1b including new data from SC 25. We see that the two parameters are highly correlated with the correlation coefficients exceeding 90% in each solar cycle. The correlation remains significant for HCMEs as well (see Fig. 1c). The HCME – SSN correlation is somewhat weaker because of the variability in the CME production of active regions.

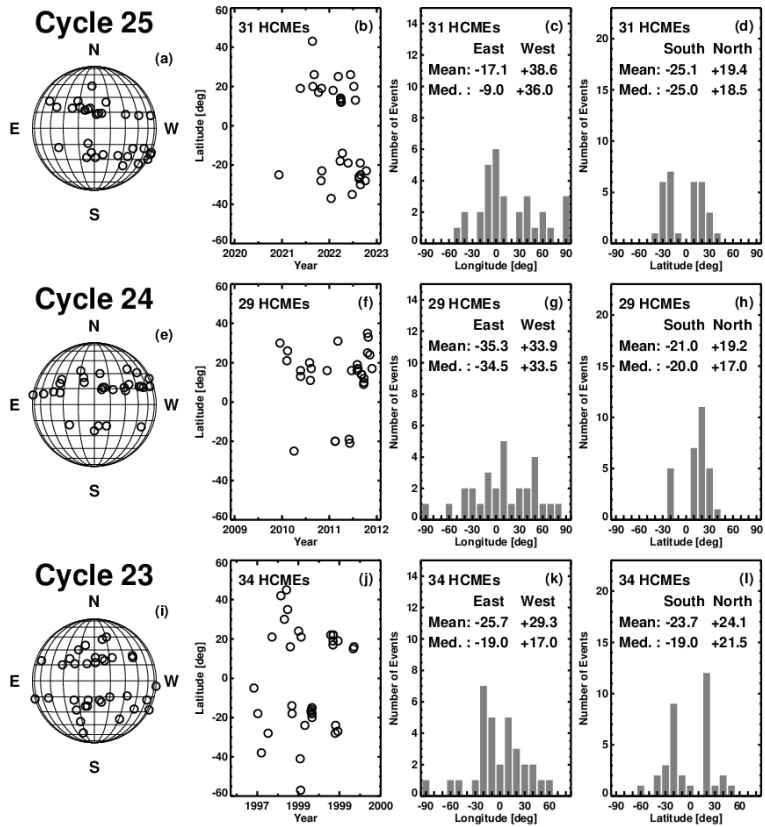

**Figure 2.** The solar source locations of HCMEs during the rise phase of the three solar cycles: heliographic coordinates (not corrected for $B_0$ angle), latitude variation with time, source longitude, and source latitude for each cycle. The hemispheric mean and median values of the source longitudes and latitudes are shown on the plots. Backside HCMEs are excluded because their source locations are unknown in cycle 23.

### 3.2 HCME source locations

One of the early findings on HCME source locations is that the number of halos originating from CMD $\geqslant 60°$ in SC 24 is twice as large as that in cycle 23. The resulting larger average CMD for SC 24 HCMEs and the higher abundance of HCMEs shown in section 3.1 has been attributed to the weak state of the heliosphere (Gopalswamy et al. 2015a; Dagnew et al. 2020). Figure 2 shows a comparison of source locations during the rise phase of the three cycles. We notice that HCMEs from CMD ⩾ 60° in SC 25 is similar to that in SC 24, but higher than that in SC 23 by a factor of ~2. The latitudes are all similar because most HCMEs originate from the active region belt. There is one more similarity between SCs 24 and 25 as to when HCMEs started appearing with respect to the cycle onset. In cycle 23, the first HCME occurred about 3 months after the cycle started on 10 May 1996. On the other hand, it took about a year for the first halos to occur in SCs 24 and 25. This is most likely due to some active regions that are highly CME productive and appeared early in the cycle.

### 3.3 CME Kinematics

Figure 3 shows the speed distributions of HCMEs during the first 37 months of the three SCs. We see that the sky-plane speeds are very similar within the typical measurement error of ~10% (Figs. 3a-c). The space speeds (three-dimensional speeds) are obtained using a cone model deprojection (Xie et al. 2004; Gopalswamy 2009). The model uses the HCME source location as input, so the space speeds are determined only for frontside halos (Figs. 3d-f). The space speeds are also similar during the rise phases. The similarity in HCME speeds between cycles 23 and 24 was reported in Gopalswamy et al. (2015a) for partial cycles. However, when the whole cycles are considered, the SC 24 speeds were significantly lower (Dagnew et al. 2020; Gopalswamy et al. 2020a). If cycle 25 is going to be weaker than SC 23, we do expect the average HCME speed in SC 25 to be smaller.

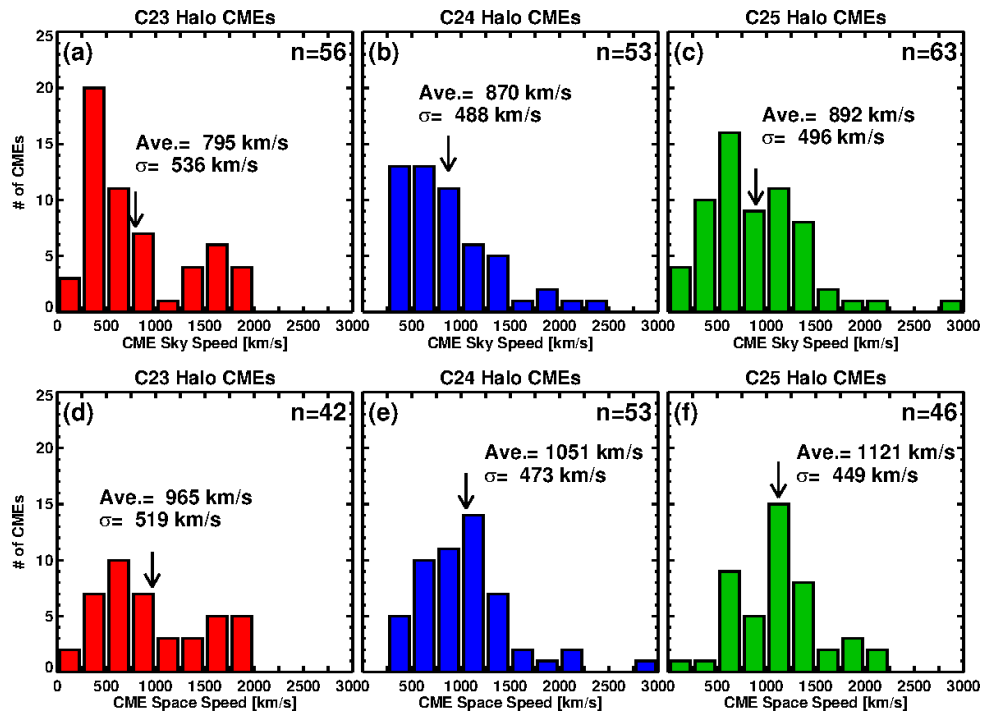

Figure 3. Sky-plane (top panels) and space speed (bottom panels) of rise-phase HCMEs of SCs 23, 24, and 25. The space speeds are obtained by a simple cone-model correction. The number of HCMEs (n) in each plot and the average speeds (with standard deviations σ) are marked on the plots.

An interesting subset of HCMEs is limb halos. Limb HCMEs are those that originate close to one the solar limbs, but the CME features appear also on the opposite limb. This happens only for very energetic CMEs (Gopalswamy et al. 2020a). During the first 37 months of each cycle, the number of limb HCMEs is very small, so it is difficult to make statistical analysis. However, the limb HCMEs do confirm the higher speeds (1396, 1259, and 1295 km/s, respectively in SCs 23, 24, and 25). The speeds of limb HCMEs are close to their true speeds because the projection effects are minimized (the sky-plane speeds can be found in https://cdaw.gsfc.nasa.gov/CME_list/halo/halo.html). The heliocentric distance at which a limb CME becomes a halo (halo height) has been found to be a good indicator of the heliospheric state (Gopalswamy et al. 2020a). The halo heights of the 7 limb halos that occurred during the rise phase of cycle 25 averages to 7.5 Rs. This is clearly lower than those in SCs 23 and 24 (see Fig. 3). The limb halos for whole cycles clearly indicated a significantly smaller halo height in SC 24 than in SC 23. This trend is not clear in Fig. 4 mainly because of low statistics. This issue needs to be revisited when the number of limb halos becomes large enough in SC 25 to properly compare with other cycles.

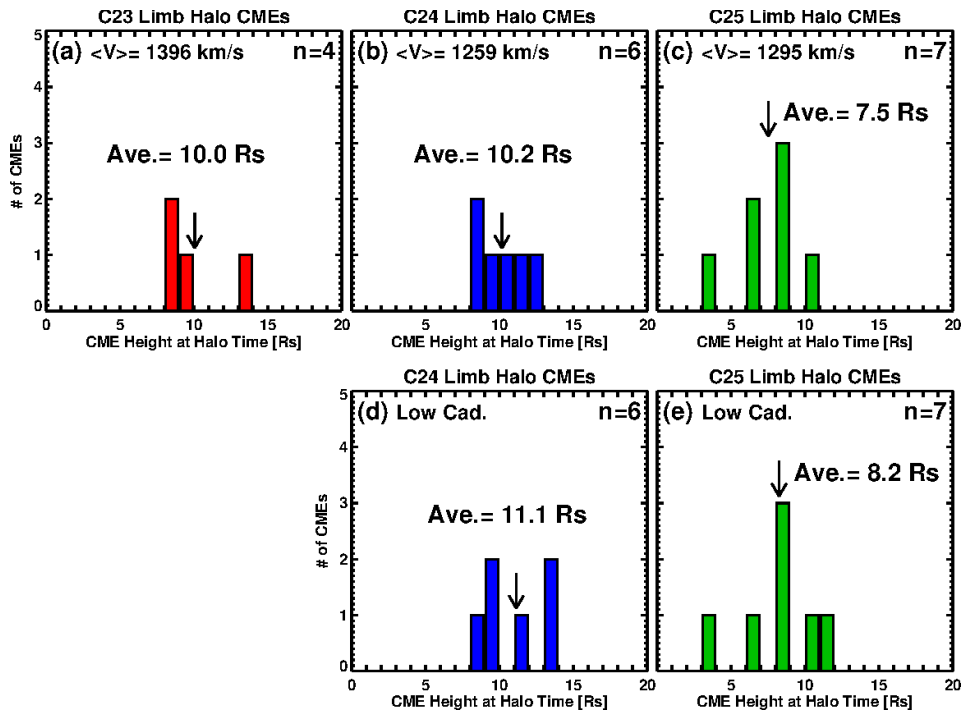

**Figure 4**. The heights at which a limb CME becomes a halo (halo heights) during the rise phase of each cycle (a-c). The top panels show the halo heights measured using the natural cadence of LASCO observations. The bottom panels show halo heights obtained by reducing the cadence of LASCO observations in cycles 24 and 25 to match that in cycle 23. The number of limb halos is small, so the statistics are poor. The average speeds of limb halos in SCs 23, 24, and 25 are marked on the plots.

### 3.4 Solar Wind Parameters in the Three Solar Cycles

The weak state of the heliosphere has been reported based on in-situ solar wind observations Gopalswamy et al. (2014; 2015b). The total pressure, magnetic field strength, density, temperature, and the Alfvén speed all declined significantly in cycle 24. The question is what happens in cycle 25. Figure 5 shows the time evolution of several solar wind parameters measured at 1-au and their extrapolations to the corona (~20 Rs). The solar wind parameters averaged over the first 37 months in each cycle are also shown on the plots. We see an interesting pattern: the cycle 25 values are in between the higher SC 23 and lower SC 24 values in every case except the Alfven speeds, which are almost the same in SCs 24 and 25. For example, the total pressure in the rise phase of SC 24 decreased by 40% with respect to SC 23. The reduction is similar (~35%) but slightly smaller in SC 25. The main result here is that the reduction in heliospheric parameters with respect to SC 23 is significant and similar in SC 25 and SC 24. This is consistent with the cycle strength predictions for SC 25 based on the precursor method: SC 25 is similar but slightly larger than SC 24, but definitely smaller than SC 23 (Nandy et al. 2021 and references therein).

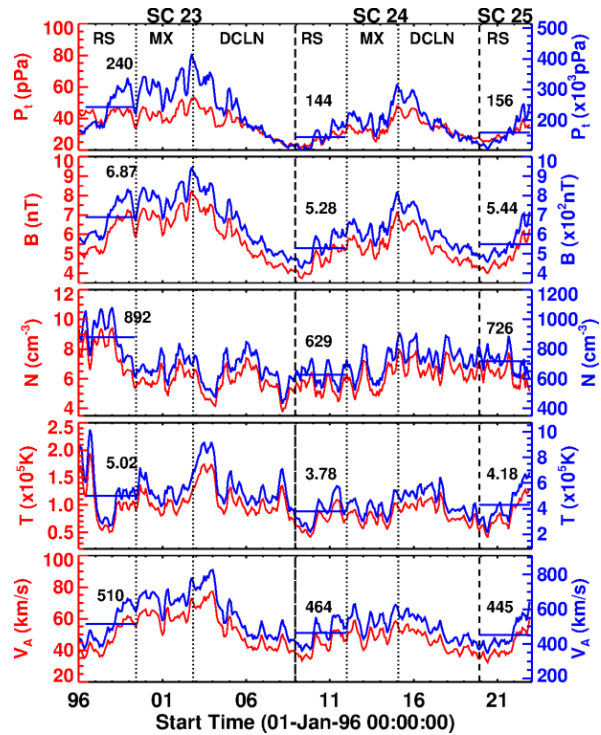

**Figure 5.** Solar Wind parameters at 1 au (red lines with left-side Y axis) from the OMNI data base: total pressure (Pt), magnetic field magnitude (B), proton density (N), proton temperature (T), and the Alfvén speed ($V_A$). The blue lines (right-side Y axis) show the same parameters extrapolated from 1 AU to the coronagraph field of view (20 Rs). The extrapolations assume that B, N, and T vary with the heliocentric distance R as $R^{-2}$, $R^{-2}$, and $R^{-0.7}$, respectively. The blue bars denote the average values of the solar wind parameters over the first 37 months, roughly corresponding to the rise (RS) phase in each solar cycle. The maximum (MX) and declining (DCLN) phases of SCs 23 and 24 are also marked for completeness. The reduction in various parameters with respect to SC 23 in SC 24 (SC 25) are: 40% (35%) in Pt, 23% (21%) in B, 30% (19%) in N, 25% (17%) in T, and 10% (13%) in $V_A$.

## 4. Discussion and Summary

There have been controversial predictions of the SC 25 strength ranging from a cycle much weaker to much stronger than SC 23. The present work utilizes the HCME abundance during the first 37 months of SCs 23-25 as an indicator of cycle strength. The use of HCMEs to assess the cycle strength stems from the fact that a weak solar cycle is accompanied by a weak heliospheric state that allows CMEs to expand more resulting in higher HCME abundance (Gopalswamy et al. 2015a; Dagnew et al. 2020). The rise phase is also appropriate for this comparison because the CME rate is tightly correlated with SSN in this phase (Gopalswamy et al. 2020b). Although we do not know the exact length of the rise phase in SC 25, we take the first 37 months to be representative of the rise phase. The main finding is that the HCME abundance in SC 25 normalized to the SSN is very similar to that in SC 24, but very different from that in SC 23. In addition, the source distribution of HCMEs is consistent with the weak state of the heliosphere because HCMEs are formed at larger CMDs than in SC 23. The weak state of the heliosphere is also evident from the diminished solar wind parameters measured at Sun-Earth L1.

We also examined other activities that distinguish between weak and strong cycles and hence corroborate the result obtained from HCMEs. In the past studies that compared the transient activities between SCs 23 and 24, two results stand out: the number of high-energy SEP events and the number of intense geomagnetic storms (Dst $\leq$ - 100 nT) are substantially lower in SC 24 than in SC 23. For example, during the first 37 months, there were 4 GLE events in SC 23 (Gopalswamy et al. 2012b). In SC 24, the first GLE did not occur until May 17, 2012, so there was no GLE in the first 37 months at all. The first GLE in SC 25 did not occur until October 28, 2021, so there was only one GLE during the first 37 months of this cycle. In SC 23, there were 20 intense geomagnetic storms. However, there were only 3 intense storms in SC 24 and one in cycle 25. Thus, both SCs 24 and 25 have a very small number of intense geomagnetic storms compared to that in SC 23. We see that the occurrence rates of GLEs and intense storms are similar in SCs 24 and 25, but much lower than the rates in SC 23. These observations clearly support the result that SC 25 is closer to SC 24, but not to SC 23.

The primary conclusions of this study can be summarized as follows:

1. The numbers of HCMEs in SCs 23, 24, and 25 during the rise phase are approximately the same, although the average SSN is substantially lower in SCs 24 and 25.
2. The CME rate – SSN correlation is strong for the general population of CMEs and HCMEs in the rise phase.
3. The HCMEs in SC 25, as in SC 24, occur over a larger CMD range than the ones in SC 23. The number of HCMEs from CMD $\geq 60°$ in SC 25 is similar to that in SC 24, but higher than that in SC 23 by a factor of ~2.
4. The average rise-phase speed of HCMEs (sky-plane and 3D) in SCs 23-25 are similar. A similar relation between SCs 23 and 24 was found before, which is applicable to SC 25 as well. However, the average speed of HCMEs ended up being smaller in SC 24 when whole cycles were compared.

5. The halo height in SC 25 is smaller than that in SCs 23 and 24, consistent with the back-reaction of the heliosphere on CMEs.
6. The solar wind total pressure (magnetic + kinetic) in SC 25 is intermediate between the high value in SC 23 and low value in SC 24.
7. The occurrence rates of GLEs and intense storms are similar in SCs 24 and 25, but much lower than the corresponding rates in SC 23.
8. In most aspects that compare SCs 23-25, SC 25 is intermediate between SCs 23 and 24, but closer to SC 24, confirming that SC 25 is similar to or only slightly stronger than SC 24, but significantly smaller than SC 23.

**Acknowledgments.** This work benefitted from NASA's open data policy in using SOHO, STEREO, and SDO data and NOAA's GOES X-ray data. SOHO is a joint project of ESA and NASA. STEREO is a mission in NASA's Solar Terrestrial Probes program. Work supported by NASA's LWS STEREO programs. HX was partially supported by NSF grant AGS-2228967. PM was partially supported by NSF grant AGS-2043131.